\documentclass[aps,twocolumn]{revtex4}
\usepackage{graphicx}
\usepackage{hyperref}

\begin{document}

%\preprint{Taras/1.7.0-pre}

\title{Molecular-beam epitaxy of (Zn,Mn)Se on Si(100)}

\author{T.~Slobodskyy}
\author{C.~R\"{u}ster}
\author{R.~Fiederling}
\author{D.~Keller}
\author{C.~Gould}
\author{W.~Ossau}
\author{G.~Schmidt}
\author{L.~W.~Molenkamp}

\affiliation{Physikalisches Institut, Universit\"{a}t W\"{u}rzburg, Am
Hubland, 97074 W\"{u}rzburg, Germany}

\date{\today}

\begin{abstract}
We have investigated the growth by molecular-beam epitaxy of the
II-VI diluted magnetic semiconductor (Zn,Mn)Se on As-passivated
Si(100) substrates. The growth start has been optimized by using
low-temperature epitaxy. Surface properties were assessed by
Nomarski and scanning electron microscopy. Optical properties of
(Zn,Mn)Se have been studied by photoluminescence and a giant
Zeeman splitting of up to 30~meV has been observed. Our
observations indicate a high crystalline quality of the epitaxial
films.
\end{abstract}



\maketitle

%\section{Introduction}
%[motivation]
The availability of the proper
material systems for spin injection, manipulation, and detection
will play an important role in further progress in the field of
spintronics.

II-VI diluted magnetic semiconductors (DMS) are
known\cite{Schmidt00} to be good candidates for effective spin
injection into a nonmagnetic semiconductor (NMS) because their
spin polarization is nearly 100\% and their conductivity is
comparable to that of typical NMS. Moreover, II-VI DMSs can be
n-type doped, thus avoiding the very fast spin precession that
limits the applicability of ferromagnetic III-V DMSs as spin
injector.

%[Why ZnMnSe]
A very promising II-VI DMS for spin injection is (Zn,Mn)Se, which
has been previously used for spin injection experiments into
GaAs\cite{Fidi00}, and ZnSe\cite{Schmidt01}.

%[Why Si]
However, these compound semiconductors are by no means the optimal
NMS materials for spin injection experiments because of their
limited spin lifetime. Silicon, because of its higher crystalline
symmetry, has a very long spin-flip length\cite{Jantsch02}.
Moreover, it is evidently in standard use in the semiconductor
industry  and therefore a very attractive material for future
spintronic devices.

It therefore seems natural to attempt spin injection from
(Zn,Mn)Se into Si. However, the surface reactivity of Si and the
lattice mismatch between Si and (Zn,Mn)Se present difficulties
that require more detailed investigation. In this paper, we
address the molecular-beam epitaxy (MBE) growth of (Zn,Mn)Se on Si
and will demonstrate that careful optimization results in high
quality epilayers.

%[previous results]
Early experiments on the growth of the parent compound ZnSe on Si
were reported in Ref.\cite{Brin92}, where using As passivation,
interfaces of reasonable quality were obtained. More recently,
Chauvet et al.\cite{Chauvet99} used the lattice match between Si
and the Zn$_{0.55}$Be$_{0.45}$Se ternary alloy, in combination
with migration enhanced epitaxy (MEE) to obtain high quality
epitaxial films. Unfortunately, (Be,Zn)Se ternary alloys with a
high Be concentration cannot be doped, making this approach
unsuitable for the growth of structures aiming at spin injection
experiments. We therefore have to resort to the growth of
large-mismatch epilayers.

%\section{Experiment}
%[surface preparation]
In order to prepare a suitable Si surface for further
heteroepitaxy, we have tried several different techniques. The
most suitable for our MBE system is a modified RCA cleaning
procedure\cite{Ish86} with subsequent hydrogen passivation.

We have found that hydrogen passivation thus obtained is stable in
air for at least 30 minutes. During this time, the wafers are
mounted on molybdenum blocks and transferred into the ultrahigh
vacuum (UHV) MBE system. Indium is used as an adhesive to mount
the substrates on the molybdenum blocks by heating it to
temperatures of about 210 $^0$C, which is below the hydrogen
desorption temperature.

%[system description]
The layers are deposited in a multichamber MBE system allowing UHV
transfer between the various growth chambers. Growth is performed
in RIBER 2300 systems using elementary sources (6N purity). Se and
As are deposited from EPI valved cracker cells and all other
elements from standard effusion cells.

%[As passivation]
After degassing at 300$^0$C for 15 min, the molybdenum blocks are
transferred to the III-V growth chamber. Immediately after the
transfer, the Reflection High Energy Electron Diffraction (RHEED)
shows (1$\times$1) patterns typical for hydrogen passivation.
After subsequent heating of the sample to approximately 730$^0$C,
the RHEED patterns change to (1$\times$2), indicating desorption
of the hydrogen passivation\cite{Eaglesham91}.

To saturate the dangling bonds and to prevent the formation of
amorphous $SiSe_{2}$\cite{Brin92} during (Zn,Mn)Se growth, an
arsenic terminated surface is prepared by cooling the sample under
arsenic flux. After this step, the Si surface is covered by a
monolayer of As. Such a surface, when tilted by 4$^0$ toward
(110), shows a (2$\times$1) RHEED pattern\cite{Kipp96}. On our
exactly (100)-oriented substrates, the patterns are (2$\times$2),
which results from a superposition of (1$\times$2) and
(2$\times$1) patterns. This can be explained by the absence of a
preferable orientation for dimer formation on an exact (100)
surface\cite {Kipp96}.

%[ZnMnSe growth]
After As passivation and cooling, the samples are transferred to
the II-VI growth chamber where (Zn,Mn)Se films are deposited.

The growth start proves critical for successful heteroepitaxial
growth of (Zn,Mn)Se on As passivated Si surfaces. The best result
is obtained using a low temperature (240$^0$C) growth start,
consisting of a Zn monolayer deposition, 10 cycles of atomic layer
epitaxy (ALE), 5 cycles of MEE and standard MBE of (Zn,Mn)Se at
300$^0$C.

This procedure yields a good separation of the  Si and Se layers
and stimulates the relaxation of the epilayer to the lattice
constant of bulk (Zn,Mn)Se. The stacking sequence of
Si-Si-As-Zn-Se-Zn-Se also allows the atoms near the interface to
be fully coordinated\cite{Brin92}.

Immediately after the start of growth, a three dimensional (3D)
growth mode is observed by RHEED. This 3D growth leads to strong
interactions between dislocations and usually yields an
improvement in structural quality. After the ALE and MEE cycles,
the growth mode stabilizes and a clear (2$\times$1) RHEED
reconstruction can be observed. During subsequent MBE, stable
(2$\times$1) RHEED patterns are observed.

%\section{Results and discussion}
%[Ellipsometry measurement]

 \begin{figure}\
 \includegraphics[width=3.375in]{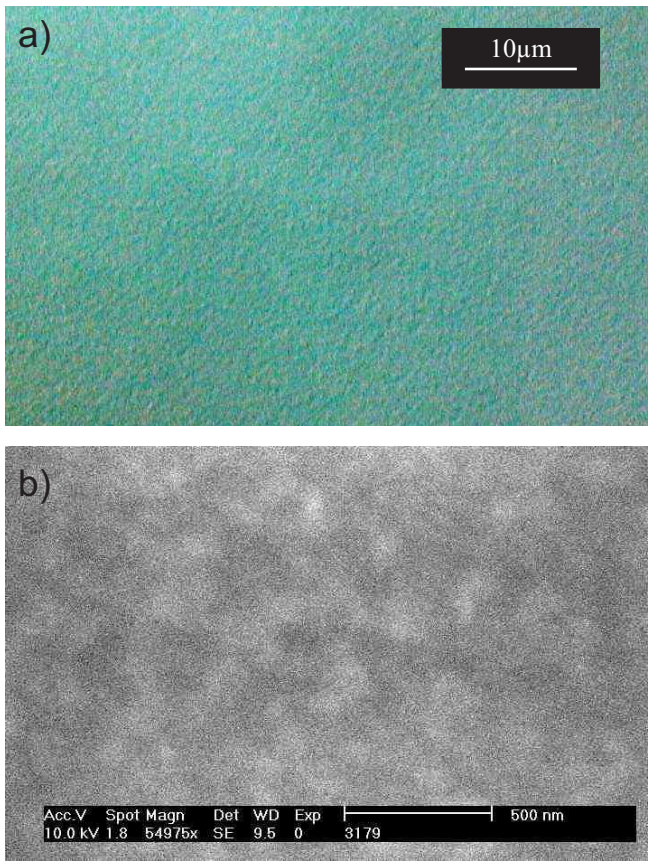}
 \caption{\label{EL+SEM} (a) Polarization differential interference
 contrast microscopy image indicating roughness of the epitaxial
 surface of (Zn,Mn)Se(100) with 4\% Mn surface. (b) Scanning electron
 microscope micrograph of the surface of a nominally identical
 sample.}
\end{figure}

Fig.~\ref{EL+SEM}(a) presents a polarization interference contrast
microscopy image of a 200~nm thick epilayer of (Zn,Mn)Se with 4\%
Mn on a Si surface oriented exactly along (100). A slight surface
roughness can be observed. When a nominally identical surface is
viewed under higher magnification using a Scanning Electron
Microscope (SEM) as in Fig.~\ref{EL+SEM}(b), a wavy surface
structure emerges. This probably originates from the 3D growth
start in combination with the lattice mismatch between the
epilayer and the substrate. The mismatch leads to strain
relaxation by local elastic deformation of the epilayer.

%[XRD measurement]

\begin{figure}\
 \includegraphics[width=3.375in]{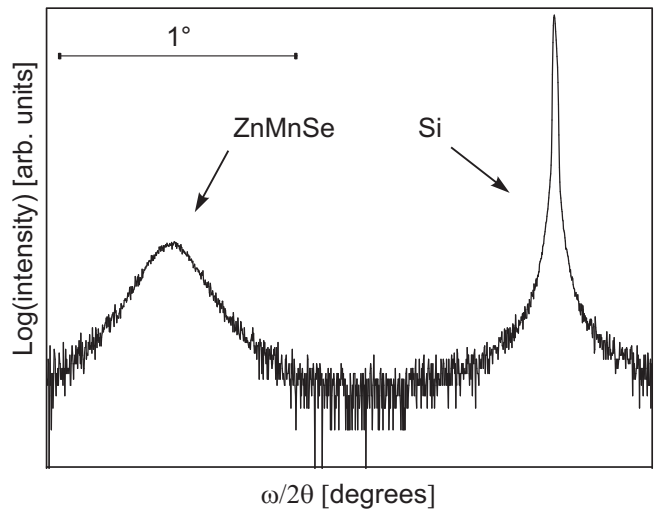}
 \caption{\label{XRD}$\omega/2\theta$ scan of a 200~nm
    (Zn,Mn)Se layer with 4\% Mn taken with a (004) reflex.}
\end{figure}

A typical High Resolution X-Ray Diffraction scan of 200~nm
(Zn,Mn)Se with 4\% Mn is shown in Fig.~\ref{XRD}. The right peak
corresponds to the silicon substrate and the left peak to the
(Zn,Mn)Se layer. The lattice mismatch of 0.23~\AA$\:$extracted
using Bragg's law from the 1.66 degree difference in peak position
agrees well with the lattice mismatch expected for a fully relaxed
(Zn,Mn)Se layer containing 4\% Mn on Si(100). An $\omega$ scan of
the (Zn,Mn)Se peak yields a Full Width at Half Maximum of about
0.4 degree, indicating an epilayer of reasonably good crystalline
quality.

%[SQUID measurement]

\begin{figure}\
 \includegraphics[width=3.375in]{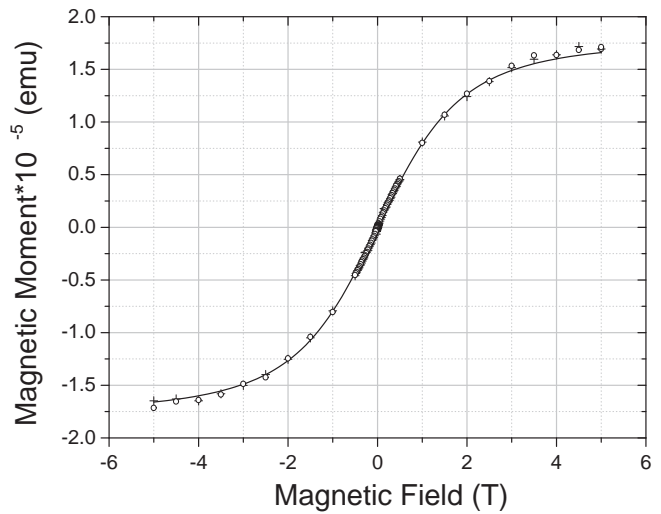}
 \caption{\label{squid} Magnetic moment of a 200~nm thick
 (Zn,Mn)Se epilayer with 4\% Mn as a function of external magnetic
 field as determined by SQUID magnetometry.}
\end{figure}

The magnetic properties of a 200~nm thick (Zn,Mn)Se epilayer with
4\% nominal Mn concentration and volume of about
10$^{6}$~$\mu$m$^{3}$ are characterized in a Quantum Design SQUID
magnetometer. A 1.8~K hysteresis loop of magnetization is
presented in Fig.~\ref{squid}. The open circles and crosses
represent the experimental data. No hysteretic behavior is
observed down to the lowest resolution of the SQUID. This places
an upper limit of one part per thousand for the fraction of Mn
atoms with ferromagnetic behavior, and excludes the possibility
that any significant portion of the Mn is present as ferromagnetic
clusters.

The data is fitted well by the solid line in the figure
corresponding to a modified Brillouin function\cite{Gaj79} that is
known to describe the functional dependence of the magnetization
on field in (Zn,Mn)Se. This fit allows us to extract an effective
temperature parameter $T_e$\cite{Gaj79} of 1.45~K, which
indicates\cite{Yu95} a Mn concentration of about 4\%. The
saturation magnetization is 0.2 Bohr magnetons per unit cell.
Using the effective Mn spin for the interacting Mn system of
(Zn,Mn)Se from Ref.~\cite{Yu95}, this corresponds to a Mn
concentration of about 3.5\%. Given the fairly large uncertainty
in the determination of the volume of the sample, we consider
these numbers to be in fair agreement. The magnetic data is
therefore entirely consistent with the incorporation of about 4\%
of Mn into the ZnSe lattice.

%[Photoluminescence measurement]

\begin{figure}\
 \includegraphics[width=3.375in]{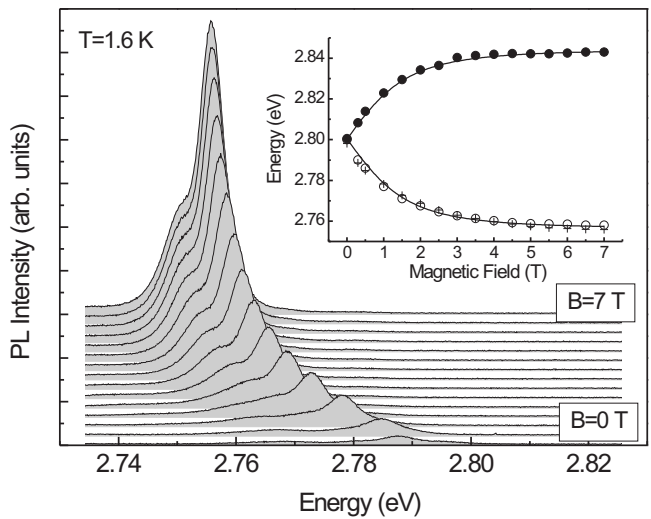}
 \caption{\label{PL} Photoluminescence spectra of 1000~nm thick
 (Zn,Mn)Se epilayer with 2\% Mn on Si(100) at various magnetic
 fields, detected in $\sigma^{+}$ - polarization. The energy dependence
 of the free exciton peak versus magnetic field is shown in the
 inset (crosses). Closed and open circles are reflectivity data
 for $\sigma^{+}$ - and $\sigma^{-}$ - polarized light,
 respectively.}
\end{figure}

The optical quality of a 1000~nm thick (Zn,Mn)Se epilayer with 2\%
Mn is examined by photoluminescence and reflectivity measurements
in a magneto-optical cryostat at 1.6~K. Fig.~\ref{PL} shows
$\sigma^{+}$ - polarized luminescence spectra of the structure for
magnetic fields of 0 to 7~T. Due to the large Zeeman splitting of
the carriers, $\sigma^{-}$ - polarized luminescence is completely
suppressed already at a very small field of about 0.3~T. At B=0,
the luminescence spectrum is dominated by a donor-bound exciton
line. With applied magnetic field, the giant Zeeman splitting
results in a suppression of the bound exciton and thus in a gain
of intensity of the free exciton line. The energy position of the
free exciton is depicted by crosses in the inset of Fig.~\ref{PL}.
The open and closed circles in this figure represent reflectivity
data, detected for $\sigma^{+}$ - and $\sigma^{-}$ - polarized
light, respectively. The Mn concentration of the epilayer can be
deduced by again fitting of the experimental value of the Zeeman
splitting with a modified Brillouin function\cite{Keller02}. The
result of the fitting procedure is indicated by lines in the inset
of Fig.~\ref{PL}. The obtained value of 3\% for the Mn
concentration is in good agreement with the growth parameters and
the magnetization measurements and indicates a good incorporation
of the Mn into the host material.

%\section{conclusions}
%[conclusions]
In conclusion we have demonstrated the high quality growth of
paramagnetic (Zn,Mn)Se epilayers on a Si(100) surface. The growth
technique and surface preparation procedure are optimized by using
As passivation and a sophisticated low temperature growth start.
Magneto-optical and magnetic characterization demonstrate that the
quality of epilayers is suitable for spin injection experiments.

%\begin{acknowledgments}
The authors would like to thank K. Brunner for useful discussions
and D. Supp for SEM measurements, as well as the DFG (SFB 410),
Darpa, and the German BMBF for financial support.
%\end{acknowledgments}

 \end{document}